\documentclass[sigconf,screen]{acmart}

\copyrightyear{2023}
\acmYear{2023}
\acmDOI{10.1145/3607199.3607200}
\acmISBN{979-8-4007-0765-0/23/10}
\acmConference[RAID '23]{The 26th International Symposium on Research in Attacks, Intrusions and Defenses}{October 16--18, 2023}{Hong Kong, Hong Kong}
\acmBooktitle{The 26th International Symposium on Research in Attacks, Intrusions and Defenses (RAID '23), October 16--18, 2023, Hong Kong, Hong Kong}

\newcounter{rq}

\newcommand{\newrq}[1]{\refstepcounter{rq}\label{rq:#1}}
\newcommand{\refrq}[1]{\textbf{\ref{rq:#1}}\xspace}

\PassOptionsToPackage{small}{caption}

\graphicspath{{./img/}}

\usepackage{amsmath,amsfonts,amssymb,amsthm}
\usepackage{siunitx}
\usepackage{graphicx}
\usepackage[english]{babel}
\usepackage{subcaption}
\usepackage{booktabs}
\usepackage{longtable}
\usepackage{multirow}
\usepackage{tabularx}
\usepackage{colortbl}
\usepackage{fancyvrb}
\usepackage{xspace}
\usepackage{upquote}
\usepackage{pdfpages}

\usepackage[inline,shortlabels]{enumitem}

\usepackage{ifxetex}
\ifxetex%
    \setcopyright{rightsretained}
    \usepackage{fontspec}
    \defaultfontfeatures{Scale=MatchLowercase}
    \defaultfontfeatures[\rmfamily]{Ligatures=TeX,Scale=1}
    \IfFontExistsTF{Libertinus Serif}{\setmainfont[]{Libertinus Serif}}{}
    \IfFontExistsTF{Noto Sans}{\setsansfont[]{Noto Sans}}{}
    \IfFontExistsTF{Iosevka}{\setmonofont[]{Iosevka}}{}
    \usepackage[cache=true,finalizecache,newfloat=true]{minted}
    \setminted{fontsize=\footnotesize}
\else
    \setcopyright{ccbysa}
    \usepackage[cache=true,frozencache,newfloat=true]{minted}
    \setminted{fontsize=\footnotesize}
\fi

\newcommand{\ffn}{F}
\newcommand{\fbinary}{B}
\newcommand{\fmutated}{\bar{B}}
\newcommand{\findices}{\mathbb{N}}
\newcommand{\flabels}{L}
\newcommand{\flstart}{\textsf{S}\xspace}
\newcommand{\flend}{\textsf{E}\xspace}
\newcommand{\flnone}{\textsf{N}\xspace}
\newcommand{\fsource}{S}
\newcommand{\ftoolchain}{T}
\newcommand{\fconfig}{C}
\newcommand{\fmuts}{M}
\newcommand{\fmut}{m}
\newcommand{\ffninf}{\hat{F}}
\newcommand{\fefp}{E^+}
\newcommand{\fefn}{E^-}
\newcommand{\fattacks}{A}

\newcommand{\mypara}[1]{\textbf{#1}.\hspace{1em}}

\begin{document}

\title{Black-box Attacks Against Neural Binary Function Detection}

\author{Joshua Bundt}
\authornote{Equal contribution.}
\affiliation{%
    \institution{Northeastern University}
  \city{}
  \country{}
}
\affiliation{%
    \institution{Army Cyber Institute}
  \city{}
  \country{}
}
\author{Michael Davinroy}
\authornotemark[1]
\affiliation{%
    \institution{Northeastern University}
  \city{}
  \country{}
}
\author{Ioannis Agadakos}
\authornote{Work done while at Northeastern University.}
\affiliation{%
    \institution{Northeastern University}
  \city{}
  \country{}
}
\affiliation{%
    \institution{Amazon}
  \city{}
  \country{}
}
\author{Alina Oprea}
\affiliation{%
    \institution{Northeastern University}
  \city{}
  \country{}
}
\author{William Robertson}
\affiliation{%
    \institution{Northeastern University}
  \city{}
  \country{}
}
\date{}

\begin{abstract}

Binary analyses based on deep neural networks~(DNNs), or \emph{neural binary analyses}~(NBAs), have become a hotly researched topic in recent years.  DNNs have been wildly successful at pushing the performance and accuracy envelopes in the natural language and image processing domains.  Thus, DNNs are highly promising for solving binary analysis problems that are hard due to a lack of complete information resulting from the lossy compilation process.  Despite this promise, it is unclear that the prevailing strategy of repurposing embeddings and model architectures originally developed for other problem domains is sound given the adversarial contexts under which binary analysis often operates.

In this paper, we empirically demonstrate that the current state of the art in neural function boundary detection is vulnerable to both inadvertent and deliberate adversarial attacks.  We proceed from the insight that current generation NBAs are built upon embeddings and model architectures intended to solve syntactic problems.  We devise a simple, reproducible, and scalable black-box methodology for exploring the space of inadvertent attacks -- instruction sequences that could be emitted by common compiler toolchains and configurations -- that exploits this syntactic design focus.  We then show that these inadvertent misclassifications can be exploited by an attacker, serving as the basis for a highly effective black-box adversarial example generation process.  We evaluate this methodology against two state-of-the-art neural function boundary detectors: XDA and DeepDi.  We conclude with an analysis of the evaluation data and recommendations for how future research might avoid succumbing to similar attacks.

\end{abstract}



\maketitle

\section{Introduction}%
\label{sec:introduction}

Binary analysis, or techniques for extracting and inferring information from code compiled to native instruction set architectures~(ISAs), is an important set of capabilities and research area in modern security. 
The field encompasses a number of distinct topics such as disassembly~\cite{ andriesse_2016_indepthanalysisdisassembly,bauman_2018_supersetdisassemblystatically,miller_2019_probabilisticdisassembly,pang_2021_sokallyou,pei_2021_xdaaccuraterobust,yu_2022_deepdilearningrelational},
function boundary detection~\cite{bao_2014_byteweightlearningrecognize,shin_2015_recognizingfunctionsbinaries,chua_2017_neuralnetscan,andriesse_2017_compileragnosticfunctiondetection,pei_2021_xdaaccuraterobust,yu_2022_deepdilearningrelational}, 
static similarity detection~\cite{haq_2019_surveybinarycode,feng_2016_scalablegraphbasedbug,xu_2017_neuralnetworkbasedgraph,massarelli_2018_safeselfattentivefunction,ding_2019_asm2vecboostingstatic,zuo_2019_neuralmachinetranslation,lee_2019_instruction2vecefficientpreprocessor,yu_2020_ordermatterssemanticaware,duan_2020_deepbindifflearningprogramwide,li_2021_palmtreelearningassembly}, 
type recovery~\cite{lee_2011_tieprincipledreverse}, 
and full decompilation~\cite{schwartz_2013_nativex86decompilation,yakdan_2015_nomoregotos,fu_2019_codaendtoendneural}.  
Each of these capabilities is in turn crucial for downstream security tasks such as malware analysis~\cite{huang_2016_mtnetmultitaskneural,raff_2018_malwaredetectioneating,al-dujaili_2018_adversarialdeeplearning,vinayakumar_2019_robustintelligentmalware} and software hardening via control-flow-integrity~(CFI) enforcement, artificial diversification, or debloating when source code is not available.

Instantiations of these capabilities in the form of deep neural networks~(DNNs) have generated substantial interest in recent years.  Neural binary analyses (NBAs) are seemingly well-matched to the problem domain, where inference is necessary due to the lossy compilation process.  Recent work has shown great promise for performing accurate disassembly~\cite{pei_2021_xdaaccuraterobust,yu_2022_deepdilearningrelational}, function boundary detection~\cite{shin_2015_recognizingfunctionsbinaries,chua_2017_neuralnetscan,pei_2021_xdaaccuraterobust,yu_2022_deepdilearningrelational}, and static binary similarity detection~\cite{feng_2016_scalablegraphbasedbug,xu_2017_neuralnetworkbasedgraph,massarelli_2018_safeselfattentivefunction,ding_2019_asm2vecboostingstatic,zuo_2019_neuralmachinetranslation,lee_2019_instruction2vecefficientpreprocessor,yu_2020_ordermatterssemanticaware,duan_2020_deepbindifflearningprogramwide,li_2021_palmtreelearningassembly} that is simultaneously more efficient than deterministic methods.

Despite this promise, questions remain as to how resilient NBAs are in practice when confronted with the incredible diversity of binary code found in the wild as well as motivated adversaries seeking to actively evade or confuse detection techniques that make use of binary analysis.  Adversarial attacks against DNNs have been intensely investigated in other problem domains~\cite{biggio_2018_wild, papernot_2018_aml_sok}, but most of these have been developed for continuous domains (e.g., images) whereas NBAs operate in a discrete domain.  Furthermore, due to the issue of problem space mapping~\cite{pierazzi_2020_intriguingpropertiesadversarial} one must develop specific black-box attacks against NBAs.

Recent work has criticized the size and scope of data used to train and evaluate NBA techniques published to date.  For instance, Kim et al.~\cite{kim_2021_revisitingbinarycode} studied NBAs that perform static similarity detection using a large dataset of programs compiled with a variety of toolchains and compiler options called BinKit.  Using this dataset and a simple baseline similarity detector called TikNib, they show that NBAs do not necessarily outperform simpler, explainable methods such as the one implemented by TikNib.  Marcelli et al.~\cite{marcelli_2022_howmachinelearning} performed a similar study also focused on static similarity detection NBAs, and show that published results do not necessarily hold when the systems-under-test are trained and evaluated on larger, more representative datasets.  Other recent work has demonstrated that DNNs used for static malware detection on binary programs are prone to adversarial attacks~\cite{lucas_2021_malwaremakeoverbreaking}, though this work relies on traditional adversarial ML techniques to either use white-box gradient descent or black-box hill climbing to find evading transformations.

In this paper, we consider the heretofore unexplored question of NBA attack resilience in the context of function boundary detection.  We focus on two exemplars of the state of the art occupying two representative points in the design space: XDA~\cite{pei_2021_xdaaccuraterobust}, which directly applies the well-known Transformer model architecture~\cite{vaswani_2017_attentionallyou} and is intended to be robust to compiler optimization level~\cite[\S4]{pei_2021_xdaaccuraterobust}, and DeepDi~\cite{yu_2022_deepdilearningrelational}, which employs a relational graph convolutional network and is explicitly advertised as intended for binary
analysis in adversarial contexts such as malware analysis -- e.g., as part of a malware analysis pipeline after dynamic analysis has been used to unpack a sample.\footnote{In their paper, the authors ``demonstrate how DeepDi is used in malware classification''~\cite[p.~2]{yu_2022_deepdilearningrelational} by evaluating their prototype on data from the Microsoft Malware Classification Challenge, comparing against MalConv~\cite{raff_2017_malconv}.}

Observing that current systems are largely based on DNN components developed to solve syntactic problems from other domains, we conjecture that these systems can be evaded using syntactic mutation.  Building on this insight, we define a simple, reproducible black-box methodology to identify misclassifying inputs to these state-of-the-art function boundary detection NBAs at scale.  Then, we demonstrate how an attacker can systematically leverage these misclassifications to either evade function detection or overwhelm a downstream analysis with false detections via at-will injection of false negatives and false positives.

From the techniques we developed, our analysis of the data leads us to several conclusions.

\begin{enumerate}[noitemsep]
    \item Sophisticated searches for adversarial examples using gradient descent are not required to significantly degrade the accuracy of NBA-based function boundary detection systems.
    \item Function boundary detection systems that build on embeddings and model architectures intended for solving syntactic problems should be viewed in a similar light as syntactic approaches for attack detection such as first-generation anti-virus and signature-based intrusion detection -- that is, with healthy skepticism.  This likely holds for other binary analysis tasks as well.
    \item It is critical that future work is evaluated on large, representative, and openly available datasets that include a range of compiler configurations as well as adversarial examples; building on existing foundations~\cite{kim_2021_revisitingbinarycode,marcelli_2022_howmachinelearning} or this work would be a good starting point.  Otherwise, it is difficult to extrapolate published evaluation results to actual operational performance.
\end{enumerate}

We note that despite these conclusions, we do not intend to completely dismiss the promise of neural binary analyses.  We discuss potential avenues for future research to mitigate the attacks found using our methodology in \S\ref{sec:discussion}.

In summary, the contributions of this paper are the following.

\begin{enumerate}[noitemsep]
    \item We propose a simple, reproducible black-box methodology for evaluating the resilience of function boundary detection NBAs to attacks at scale.
    \item We demonstrate the susceptibility of the current state of the art, represented by XDA~\cite{pei_2021_xdaaccuraterobust} and DeepDi~\cite{yu_2022_deepdilearningrelational}, to producing overwhelming false negatives and false positives to downstream binary analyses.
    \item We discuss and synthesize conclusions from an analysis of the evaluation data, and suggest several paths forward to mitigate similar attacks against neural binary analysis.
\end{enumerate}

The source code and datasets are available at~\url{https://osf.io/bcdxq/}.

\section{Problem Statement and Motivation}%
\label{sec:motivation}

\subsection{Binary Analysis}%
\label{sub:binary_analysis}

The term ``binary analysis'' encompasses a wide range of techniques that all attempt to extract information from programs that have been compiled to a native instruction set architecture~(ISA).  These techniques range from fundamental analyses such as disassembly~\cite{pang_2021_sokallyou,pei_2021_xdaaccuraterobust,yu_2022_deepdilearningrelational} and function boundary
detection~\cite{bao_2014_byteweightlearningrecognize,chua_2017_neuralnetscan,pei_2021_xdaaccuraterobust,yu_2022_deepdilearningrelational} to downstream tasks that build on prior analyses such as static similarity detection~\cite{haq_2019_surveybinarycode,feng_2016_scalablegraphbasedbug,zuo_2019_neuralmachinetranslation,yu_2020_ordermatterssemanticaware,li_2021_palmtreelearningassembly}, type recovery~\cite{lee_2011_tieprincipledreverse}, malware
detection~\cite{huang_2016_mtnetmultitaskneural,raff_2018_malwaredetectioneating,al-dujaili_2018_adversarialdeeplearning,vinayakumar_2019_robustintelligentmalware}, and full decompilation~\cite{schwartz_2013_nativex86decompilation,yakdan_2015_nomoregotos,fu_2019_codaendtoendneural}.  Designing accurate and efficient binary analyses is substantially more difficult than for source code due to the inherently lossy compilation process.  That is, compiler toolchains discard much of the higher-level
abstractions present in source code when lowering to an ISA.  Thus, binary analyses must operate with incomplete information and are virtually always unsound.  Compounding this difficulty is that binary analyses are often, though not always, performed under a strong threat model~\cite{ren_2021_bintuner} in which active adversaries attempt to evade or otherwise confuse those analyses.

While binary analyses have traditionally employed deterministic methods, the lack of source code naturally suggests inference methods as a promising approach for improving both accuracy and performance.  In that light, it should come as no surprise that deep neural networks~(DNNs) have come to the fore as a basis for binary analysis research.  Table~\ref{tab:nba_techniques} presents an overview of recent work in this vein, to which we refer hereinafter as \emph{neural binary analyses} or \emph{NBAs}.

\begin{table*}[t]
    \footnotesize
    \centering
    \begin{tabular}{llllc}
        \toprule
        \textbf{System} & \textbf{Input} & \textbf{Embedding} & \textbf{Architecture} & \textbf{Capabilities} \\
        \midrule
        BiRNN~\cite{shin_2015_recognizingfunctionsbinaries} & Bytes & One-hot encoding & RNN & F \\
        MtNet~\cite{huang_2016_mtnetmultitaskneural} & API calls, memory objects & Bit vector & Feedforward & M \\
        Eklayva~\cite{chua_2017_neuralnetscan} & Disassembly text & word2vec~\cite{mikolov_2013_efficientestimationword} & RNN & F \\
        Gemini~\cite{xu_2017_neuralnetworkbasedgraph} & ACFG & structure2vec~\cite{dai_2016_discriminativeembeddingslatent} & Siamese & S \\
        Sleipnir~\cite{al-dujaili_2018_adversarialdeeplearning} & Windows API invocations & Bit vector & Feedforward & M \\
        SAFE~\cite{massarelli_2018_safeselfattentivefunction} & Disassembly text & word2vec~\cite{mikolov_2013_efficientestimationword} & BiRNN, Siamese & S \\
        asm2vec~\cite{ding_2019_asm2vecboostingstatic} & Disassembly text & PV-DM~\cite{le_2014_distributedrepresentationssentences} & LSTM & S \\
        Coda~\cite{fu_2019_codaendtoendneural} & Disassembly text & AST & Tree-LSTM~\cite{tai_2015_improvedsemanticrepresentations} & R \\
        InnerEye~\cite{zuo_2019_neuralmachinetranslation} & Disassembly text & word2vec~\cite{mikolov_2013_efficientestimationword} & LSTM & S \\
        Instruction2Vec~\cite{lee_2019_instruction2vecefficientpreprocessor} & Disassembly text & word2vec~\cite{mikolov_2013_efficientestimationword} & CNN & S \\
        Order Matters~\cite{yu_2020_ordermatterssemanticaware} & Disassembly text & BERT~\cite{devlin_2019_bertpretrainingdeep}, CNN & MPNN~\cite{gilmer_2017_neuralmessagepassing}, CNN & S \\
        DeepVSA~\cite{guo_2019_deepvsafacilitatingvalueset} & One-hot encoding & Context vectors & LSTM & V \\
        DeepImgMalDetect~\cite{vinayakumar_2019_robustintelligentmalware} & Pixels & --- & RNN & M \\
        DeepBinDiff~\cite{duan_2020_deepbindifflearningprogramwide} & Disassembly text & word2vec~\cite{mikolov_2013_efficientestimationword} & ANN & S \\
        XDA~\cite{pei_2021_xdaaccuraterobust} & Bytes & One-hot encoding & Transformer~\cite{vaswani_2017_attentionallyou} & D, F \\
        PalmTree~\cite{li_2021_palmtreelearningassembly} & Disassembly text & BERT~\cite{devlin_2019_bertpretrainingdeep} &  --- & F, S, V \\
        Codee~\cite{yang_2021_codeetensorembedding} & Disassembly text & word2vec~\cite{mikolov_2013_efficientestimationword}, node2vec~\cite{grover_2016_node2vecscalablefeature} & --- & S \\
        DeepDi~\cite{yu_2022_deepdilearningrelational} & Instruction metadata & RNN & Relational-GCN~\cite{schlichtkrull_2018_modelingrelationaldata} & D, F \\
        \bottomrule
    \end{tabular}
    \caption{Summary comparison of various neural binary analysis systems. Note that all of these systems are at least in part built on embeddings and model architectures developed to solved problems in the NLP or image processing domains. \emph{Capabilities:} D = disassembly, F = function identification, V = value set analysis, S = similarity, R = decompilation, M = malware detection or classification.}%
    \label{tab:nba_techniques}
\end{table*}

Each entry in Table~\ref{tab:nba_techniques} lists the input, embedding, model architecture, and the binary analyses implemented.  An embedding is simply a procedure for mapping input data to a representation on which that model performs training and inference.  Common choices of embeddings are one-hot encoding of byte sequences, or text embeddings such as word2vec~\cite{mikolov_2013_efficientestimationword} applied to the token stream produced by a disassembler.  The model architecture, on the other hand, is the neural network proper; that is, the set of layers, interconnections, and weights responsible for inference.  It is common for NBAs to repurpose model architectures developed for natural language or image analysis tasks; examples include recurrent neural networks~(RNNs), convolutional neural networks~(CNNs), and  the Transformer architecture~\cite{vaswani_2017_attentionallyou}.

\mypara{Motivation}%
Recent prior work has studied the accuracy of NBAs for static similarity detection~\cite{kim_2021_revisitingbinarycode,marcelli_2022_howmachinelearning} and malware detection~\cite{lucas_2021_malwaremakeoverbreaking}.  However, to the best of our knowledge, the question of whether NBAs are a suitable solution for function boundary detection has not been definitively studied.  This paper attempts to answer this question, and thus focuses specifically on prominent systems targeting the function boundary detection task.

\subsection{Function Boundary Detection}%
\label{sub:function_boundary_detection}

Function boundary detection is a fundamental binary analysis that typically occurs directly after, or even in tandem with, disassembly~\cite{pang_2021_sokallyou}.  Identifying functions is crucial for many downstream tasks.  For instance, most static similarity algorithms consider pairs of functions when computing distances.  Functions are also important inputs to recursive descent disassemblers as starting points for recursive disassembly or as possible callees of indirect call sites.

If function detection is performed as part of a manual process -- e.g., interactive reverse engineering provided by tools like IDA Pro~\cite{guilfanov_2022_idapro}, Ghidra~\cite{nsa_2019_ghidra}, or Binary Ninja~\cite{vector35_2016_binaryninja} -- excessive false positives could lead to user fatigue and, in turn, an unusable tool~\cite{axelsson_2000_baseratefallacydifficulty}.  False negatives, on the other hand, are perhaps even more concerning since failing to identify functions could directly lead to evasion opportunities for attackers that aspire to elude detection.

Function boundary misclassifications can also have a large impact on the accuracy and utility of an automated analysis pipeline.  For instance, it is common to combine successive rounds of static and dynamic analysis to, e.g., first unpack a malware sample in a sandbox so that an efficient static analysis can be performed on an unobfuscated dropped or in-memory binary~\cite{yongwong_2021_lookpracticemalware}.  False negative function detections in this scenario could again lead to detection ``blind spots,'' while false positives could degrade the efficiency or accuracy of downstream analyses.

More formally, we can think of a function boundary detection NBA as a procedure that learns a mapping from bytes or instructions in a binary, depending on the embedding, to one of three labels: \flstart for function entry points, \flend for function exit points, and \flnone for all other points.  Let \(\fbinary\) be the set of binary inputs and \(\findices\) be the set of possible byte or instruction indices in each binary.  We can then denote this mapping as
\begin{align}
    \ffn : \fbinary \times \findices \mapsto \flabels = \left\{ \flstart, \flend, \flnone \right\}. \label{eq:ffn}
\end{align}
Early work in the NBA space heavily borrowed from DNNs built to tackle natural language processing~(NLP) problems.  The first system to adopt this approach was BiRNN~\cite{shin_2015_recognizingfunctionsbinaries}, which treated byte sequences comprising binaries as tokens in a language.  BiRNN converts each input byte into \(\mathbb{R}^{256}\) vectors using a one-hot encoding, where a byte's value is indicated by the position of the single 1 in a vector.  Encoded bytes are then fed to a bi-directional RNN, where the use of two RNNs allows for prediction of a byte label using both preceding and succeeding bytes as context.

XDA~\cite{pei_2021_xdaaccuraterobust} built upon BiRNN's approach to function boundary detection by adapting another powerful model architecture from the NLP literature: Transformer~\cite{vaswani_2017_attentionallyou}.  Transformer pioneered the concept of \emph{self-attention}, where an attention layer allows the model to process sequential data out of order.  This allows Transformer-based models to flexibly learn and infer meaning from context as well as parallelize better than prior architectures like RNN, LSTM, and GRU.  Transformer-based models such as BERT~\cite{devlin_2019_bertpretrainingdeep} (Bidirectional Encoder Representations from Transformers) and the GPT family~\cite{brown_2020_languagemodelsare} (Generative Pre-Trained Transformer) represent the state of the art in NLP model architectures.

XDA's implementation~\cite{kexinpei_2021_xda} directly applies a popular implementation of BERT called RoBERTa (provided by Facebook's Fairseq~\cite{ott_2019_fairseq} library) to the binary disassembly and function boundary detection tasks.  Binaries are processed in 512-byte chunks, and a one-hot encoding is used to produce \(\mathbb{R}^{256}\) vectors to be processed by the network.  In addition to byte values, the input vocabulary defines five additional tokens representing \emph{padding},
\emph{start-of-sequence}, \emph{end-of-sequence}, \emph{unknown}, and \emph{mask} (not all are used by XDA).  In the first of two phases, the model is pre-trained using masked language modeling~(MLM), which essentially teaches the model to predict byte values given surrounding context.  The resulting model is then fine-tuned in the second phase to transfer the knowledge learned in the first phase to a particular binary analysis task such as function boundary detection.

DeepDi~\cite{yu_2022_deepdilearningrelational} is a state-of-the-art example of an NBA-based disassembly and function boundary detection system.\footnote{DeepDi only recovers function entry points, and so is more accurately called a function start detection system.}  While DeepDi follows in the tradition of BiRNN and XDA by building upon existing model architectures, in this case, R-GCN~\cite{schlichtkrull_2018_modelingrelationaldata} (Relational Graph Convolutional Model), it improves on prior work in several ways.  First, it eschews the use of deep learning altogether for the initial disassembly step, choosing instead to rely on superset disassembly~\cite{bauman_2018_supersetdisassemblystatically} to recover all possible instructions contained in an input binary.  The instruction superset, in the form of 
4-tuples of \(\langle \textsf{opcode}, \textsf{mod}\_\textsf{rm}, \textsf{scale}\_\textsf{index}, \textsf{rex}\_\textsf{prefix} \rangle\), is then converted into a fixed-dimension embedding using a learned embedding layer.  Each embedding is concatenated with the following two instruction embeddings which is fed to an RNN to arrive at a final instruction representation.  These representations then serve as input to the R-GCN, which models various relationships between instructions using an Instruction Flow Graph~(IFG) in order to weed out invalid instructions from the superset and retain only the ``true'' disassembly.

To identify function entry points, DeepDi first collects a set of candidate entry points by applying a set of heuristics to instructions identified as valid from the superset.  Each candidate instruction is packed with the three preceding and three succeeding instructions and then fed to the entry point recovery model.  This model consists of an embedding layer, a GRU layer, and a two-layer perceptron classifier.  The authors of DeepDi note that while the model achieved an average F1 score of \num{98.6}\% on the function start detection task in their evaluation, their heuristics-based approach ``will miss tail jumps and functions with unseen prologues~\cite[p.~7]{yu_2022_deepdilearningrelational}.''

\subsection{Semantics, or Merely Syntax?}%
\label{sub:semantics}

While systems like XDA make repeated reference to ``learning semantics,'' these representations do not encode the semantic outcome of the input when executed on a system. We conjecture this limitation is due to the approach of being trained using only disassembled instructions or raw sequences of bytes extracted from binaries, as correspondence is limited to patterns of bytes or textual tokens presented during training.  Absent of semantic meaning, code isomorphisms that syntactically appear drastically different might well not be detected as semantically equivalent.

To illustrate, Listing~\ref{lst:code_optimizations} presents a na\"{i}ve addition function and its compilation to x86\_64 assembly using two commonly available optimization levels: \texttt{O0} and \texttt{O3}.  The resulting code, while semantically equivalent, has radically different syntactic forms, and systems relying only on detecting sequences of bytes or instructions would fail to identify the optimized version if they have not encountered a similar example during training.  While an argument can be made for generating comprehensive datasets that contain both versions (and indeed virtually all current methods do try to include these common compiler optimizations), we argue that such an approach cannot scale to include every possible combination of all available compiler flags.  In essence, this places a hard constraint on what is possible for NBA models to learn in the absence of semantic information.

\begin{listing}[t]
    \rule{\linewidth}{0.5pt}
\begin{minted}{c}
int add(int a, int b) {
    if (a == 0)
        return b;
    else if (b == 0)
        return a;
    return a + b;
}
\end{minted}
    \rule{\linewidth}{0.5pt}
\begin{minted}{nasm}
add_O0:
    push  rbp                   ; save caller fp
    mov   rbp, rsp              ; set fp
    mov   dword [rbp-0x4], edi  ; get arg1
    mov   dword [rbp-0x8], esi  ; get arg2
    cmp   dword [rbp-0x4], 0x0  ; compare arg1 to 0
    jne   .check_arg2
    mov   eax, dword [rbp-0x8]  ; return arg2
    jmp   .return
.check_arg2:
    cmp   dword [rbp-0x8], 0x0  ; compare arg2 to 0
    jne   .do_add
    mov   eax, dword [rbp-0x4]  ; return arg1
    jmp   .return
.do_add:
    mov   edx, dword [rbp-0x4]
    mov   eax, dword [rbp-0x8]
    add   eax, edx              ; return arg1 + arg2
.return:
    pop   rbp                   ; restore caller fp
    ret                         ; return to caller
\end{minted}
    \rule{\linewidth}{0.5pt}
\begin{minted}{nasm}
add_O3:
    lea    eax, [rdi+rsi*1]     ; return arg1 + arg2
    ret                         ; return to caller
\end{minted}
    \rule{\linewidth}{0.5pt}
    \caption{Compiler optimizations can have a drastic effect on program representation in compiled code.  Relying only on sequences of bytes or textual tokens absent of semantic information limits detection only to known syntactic patterns.}
    \label{lst:code_optimizations}
\end{listing}

In the remainder of this paper, we build upon this insight to demonstrate systematic attacks against neural binary analyses for function boundary detection.

\section{Attacking Neural Function Boundary Detection}%
\label{sec:methodology}

\begin{figure*}[t]
    \centering
    \includegraphics[width=\textwidth]{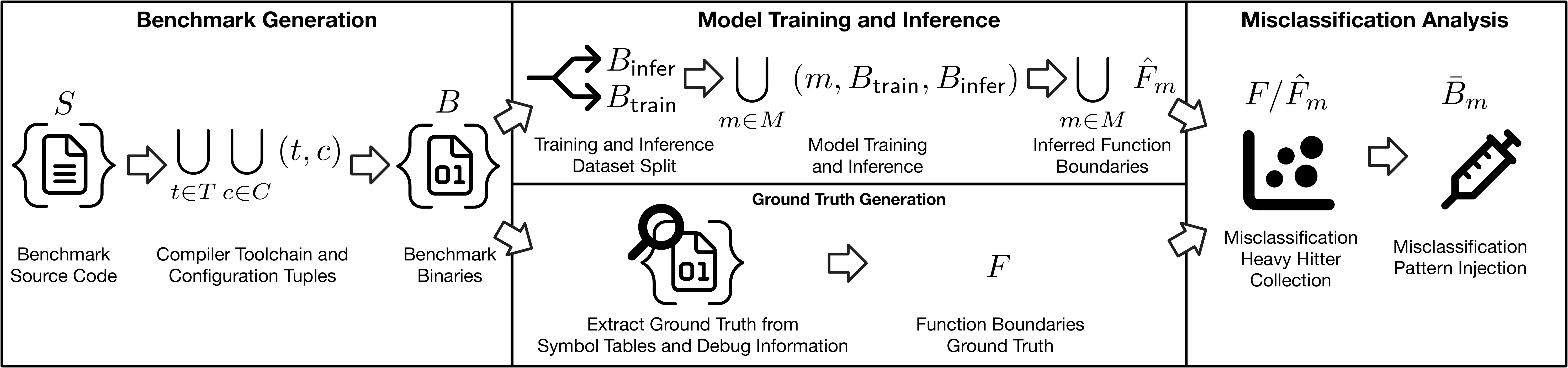}
    \caption{Overview of the NBA function boundary detection vulnerability search procedure.  In the first phase, benchmark source code \(\fsource\) is compiled by an array of compiler toolchains and configurations \(\langle \ftoolchain, \fconfig \rangle\) resulting in a benchmark binary corpus \(\fbinary\).  Function boundary ground truth \(\ffn\) is extracted from \(\fbinary\).  In parallel, the set of models-under-test \(M\) is trained and evaluated on one or more training and inference
    splits of \(\fbinary\).  Finally, misclassifications in the form of false positives and negatives are collected by comparing \(\ffn,\bigcup_{\fmut \in \fmuts} \ffninf_m\).  Heavy hitters are identified and injected into \(B\) for attack evaluation.}%
    \label{fig:overview}
\end{figure*}

In our evaluation of neural function boundary detection, we focus on \emph{black-box attacks}.  These attacks are so-named since no information about the model-under-test (MUT) such as its internal weights or structure are assumed.  Black-box attacks are advantageous because they do not require a deep understanding of MUTs; instead, only the ability to issue queries and observe results is needed.  However, as the search process is unguided by model information, black-box techniques can fail to discover latent vulnerabilities that a white-box adversarial search such as projected gradient descent~(PGD)~\cite{madry_2019_deeplearningmodels} might otherwise uncover.  In this sense, the results of our methodology should be considered a lower bound on the vulnerability of MUTs to which it is applied.

We define a general black-box vulnerability search procedure with the goal of uncovering and exploiting function boundary misclassifications when performing inference on binary programs.  The search proceeds in several phases:
\begin{enumerate*}[(i)]
    \item input generation,
    \item ground truth generation,
    \item training and inference, and
    \item misclassification analysis.
\end{enumerate*}

\mypara{Input Generation}\label{par:input_generation}%
In the first phase, we gather a corpus of benchmark program source code \(\fsource\).  Each benchmark is compiled by an array of compiler toolchains \(\ftoolchain\), each of which is equipped with an attack configuration \(\fconfig\) consisting of a set of compiler flags and code transformations.  Given \(n = \left| \ftoolchain \right|\) compilers, we obtain a benchmark binary corpus \(\fbinary\) consisting of \(n\) separate compilations of \(\fsource\) with each toolchain and attack configuration tuple \(\langle \ftoolchain, \fconfig \rangle\).

\mypara{Ground Truth Generation}\label{par:ground_truth_generation}%
Each configuration ensures that debugging information is generated while simultaneously preventing compiled binaries from being stripped of symbol information.  Thus, we can post-process each binary and use these sources of information to construct a ground truth mapping \(\ffn\)~\eqref{eq:ffn}; that is, a function that labels each byte of code in each binary as to whether it is a function start, a function end, or neither.

\mypara{MUT Training and Inference}\label{par:mut_training_inference}%
In parallel, we split \(\fbinary\) into training and inference sets.  The MUTs \(\fmuts\) are individually trained and evaluated on these sets.  The result for each MUT \(\fmut \in \fmuts\) is an \emph{inferred} mapping \(\ffninf_\fmut\).  Since we extracted a ground truth labeling \(\ffn\) in the previous phase, we can directly compare \(\ffninf_\fmut\) and \(\ffn\) to identify misclassifications in the form of false positives \(\fefp_\fmut\) and false negatives \(\fefn_\fmut\) where
\begin{align}
    \fefp_\fmut = \left\{  b, i \right\} \; \mathrm{ s.t. } \; &\forall \; b \in \fbinary, i \in \left| b \right| \; \ffn(b, i) = \flnone \land \ffninf_\fmut(b, i) \in \left\{  \flstart, \flend \right\} \\
    \fefn_\fmut = \left\{ b, i \right\} \; \mathrm{ s.t. } \; &\forall \; b \in \fbinary, i \in \left| b \right| \; \ffn(b, i) \in \left\{ \flstart, \flend \right\} \land \ffninf_\fmut(b, i) = \flnone
\end{align}

\mypara{Misclassification Analysis}\label{par:misclassification_analysis}%
In the last phase, for each MUT we process its misclassification sets \(\fefp_\fmut,\fefn_\fmut\) to identify \emph{attack inputs} \(\fattacks_\fmut\) that can reliably produce function boundary misclassifications in arbitrary binary programs.  To do so, we rank-order misclassifications for each model by highest incidence to lowest.  The ranked attack inputs then serve as seeds for an adversarial search, where they are each injected in turn into targeted functions of \(\fbinary\) to produce a mutated corpus \(\fmutated_m\).  A separate attack validation round is then carried out by having the MUT \(\fmut\) perform inference on \(\fmutated_\fmut\) to confirm that the intended misclassifications are replicated in the targeted functions.

\subsection{Attack Techniques and Threat Models}%
\label{sub:threat_model}

The vulnerability search procedure relies upon collecting a set of attack techniques in the form of compiler flags and code transformations.  Each of these techniques specifically targets function \emph{prologues} and \emph{epilogues}.  A function prologue is responsible for
\begin{enumerate*}[(i)]
    \item saving the contents of any registers that it uses and that a caller is responsible for preserving under a given calling convention; and,
    \item allocating space on the current thread's stack for any local variables the function uses.
\end{enumerate*}
An epilogue, on the other hand, is responsible for reversing the effects of the prologue as well as (optionally) returning a value to the caller.  Our attack techniques modify function prologues and epilogues because neural function boundary detection models focus on bytes or instructions that comprise (or are adjacent to) these code regions.

However, while each attack technique targets the same code regions, not all techniques are created equally.  Some attacks are \emph{inadvertent}; that is, an unintended deficiency of the data representation, network architecture, or training set causes the model to misclassify a benign input during inference.  Other attacks, however, are inherently \emph{adversarial}.  In this case, an adversary intentionally transforms their input to actively attack the MUT.  In this case, the only restriction is that the transformation must preserve the intended functionality of the attack.

\mypara{Inadvertent Threat Model}%
Inadvertent attacks represent a weak threat model, in that there is no active adversary and, instead, the MUT or training set is suboptimal with respect to a ``naturally-occurring'' binary that has been emitted by a standard compiler toolchain on benign code.  While this threat model is weaker, Ren et al.\ show that in practice ``adversaries explore non-default compiler settings to amplify malware differences''~\cite{ren_2021_bintuner}.

\mypara{Adversarial Threat Model}%
In this threat model, no assumptions are made about how the binary was produced.  Binaries can be obfuscated or encrypted, and in such cases must be unpacked in a malware sandbox prior to use of a static analysis on an unobfuscated dropped or in-memory executable image, as is common industry practice~\cite{yongwong_2021_lookpracticemalware}.

We classify the attack techniques employed by the search procedure according to whether they are inadvertent or adversarial.  The criterion we use for this classification is whether or not the code resulting from applying a technique can be emitted by an unmodified compiler toolchain given a legal configuration.

An overview of the vulnerability search procedure is shown in Figure~\ref{fig:overview}.  In the following, we describe each inadvertent and adversarial attack technique.

\subsection{Inadvertent Attacks}%
\label{sub:inadvertent_attacks}

Inadvertent attacks result from misclassified binary code emitted by a benign compiler toolchain under any possible configuration.  However, systematically exploring the entire space of possible compiler configurations in terms of combinations of compiler flags is daunting, to put it lightly.  To illustrate, clang~v13.1.6 (arm64-apple-darwin21.4.0) advertises \num{1013} distinct command-line options in its default help message when invoked using \texttt{clang --help}.  Thus, if we denote the set of possible options as \(C\), a rough estimate of number of possible combinations is \( \left| \mathcal{P}\left( C \right) \right| = 2^{1013} \).\footnote{This is a loose estimate.  It is likely that some combinations are invalid, which would lead to an overcount.  However, some options are not boolean flags but rather take a value as an argument, which would lead to an undercount.}  An exhaustive exploration of this space is clearly intractable in practice, and so we use domain knowledge to select a small number of compiler options that we conjecture will have an effect on function boundary prediction.  We describe each of these classes of options below.\footnote{We also note that we do not claim these classes as exhaustive.  Indeed, we are aware of other compiler configurations and code transformations that would be interesting to explore.  Unfortunately, due to time constraints we have not yet fully evaluated them and so elide them here.}

\mypara{Stack Protector}\label{par:stack_protector}%
``Stack protector'' is gcc and clang's modern name for canary/cookie-based anti-stack smashing defenses~\cite{cowan_1998_stackguardautomaticadaptive}.  This defense injects an unpredictable guard value onto the stack as part of the function prologue.  In an epilogue, the injected copy of the guard is compared to a global copy.  If the values do not match, then a stack smashing attack is assumed to have occurred and the program is terminated before the attacker can gain control of execution via, e.g., a corrupted return address.  Otherwise, execution continues as normal.

The defense relies upon several assumptions -- for instance, that the guard value contains carries sufficient entropy to make guessing infeasible, that the guard value is not leaked to the adversary, that the global copy of the guard cannot be modified by the adversary, and that all stack-allocated data that can be leveraged to hijack code execution is protected by the stack guard.  The defense can in fact take several forms depending on the compiler version and particular flags used, such as: whether all functions are protected by a guard or rather only those that allocate a buffer on the stack; whether some or all stack variables are protected by the guard, which can involve variable reordering; and, the offset of the global guard copy.

Our inclusion of these compiler options is based on the observation that stack guard injection and verification requires modifications to function prologues and epilogues.  NBAs that rely on particular byte or instruction sequences comprising prologues and epilogues for function boundary detection might thus be confused by these modifications.  Listing~\ref{lst:stack_protector}~(\S\ref{sec:inadvertent_evasion_effects}) illustrates a typical example.

\mypara{Stack Clash Protection}\label{par:stack_clash_protection}%
Stack clash vulnerabilities arise when an attacker is able to grow either the stack or another memory region such that the two memory segments overlap~\cite{qualys_2017_qualyssecurityadvisory}.  While OS kernels such as Linux can inject a guard page to separate the stack from other regions, prior research has shown that guard pages are nevertheless circumventable.  Thus, modern compilers include options to enable stack clash mitigations in emitted code.  The most popular form of this mitigation centers on breaking large stack allocations into page-sized chunks, and either implicitly or explicitly probing each chunk to ensure that it has not clashed with an existing memory allocation~\cite{law_2020_stackclashmitigation}.

The impetus for our inclusion of stack clash protector compiler options as an attack technique is the modified allocation pattern for large stack buffers in function prologues and the requirement for explicit probe injection if the compiler deems it necessary.  Listing~\ref{lst:stack_clash}~(\S\ref{sec:inadvertent_evasion_effects}) illustrates one form of these modifications to function prologues (epilogues are not affected in this case).

\mypara{Control Flow Integrity}\label{par:cfi}%
Control flow integrity~(CFI) is a general software hardening approach based on the principle that code must execute control transfers if and only if those transfers were intended by the programmer~\cite{abadi_2009_controlflowintegrityprinciples}.  Forward-edge CFI in particular has become a standard feature of production compilers like clang and gcc~\cite{tice_2014_enforcingforwardedgecontrolflow}, efficiently protecting indirect calls and jumps through computed pointers of various forms.  Architectural support for a weak form of return-edge CFI has also become available in recent x86/x86-64 processor generations in the form of Intel CET~\cite{shanbhogue_2019_securityanalysisprocessor}.  While forward-edge CFI checks such as IFCC~\cite{tice_2014_enforcingforwardedgecontrolflow} are typically inserted at call sites, Intel CET enforcement depends on instrumenting valid indirect branch targets with special instructions (\texttt{endbr32}, \texttt{endbr64}) as well as ensuring that these instructions do not accidentally appear anywhere else in an executable memory region.  Since this instrumentation can result in modified function prologues, we include CFI enforcement options as a separate attack technique.  Listing~\ref{lst:cfi}~(\S\ref{sec:inadvertent_evasion_effects}) illustrates function prologue modifications resulting from Intel CET and indirect branch tracking enforcement.

\mypara{SafeStack}\label{par:safestack}%
SafeStack is another stack-based buffer overflow defense developed as part of code-pointer integrity (CPI)~\cite{kuznetsov_2014_codepointerintegrity} that relies upon separating stacks into safe and unsafe stacks.  Security-relevant data such as return addresses, register spills, and local variables are stored on the safe stack.  Accesses to the safe stack are always checked via runtime instrumentation for safety.  All other stack-allocated data is stored on the unsafe stack, ensuring that buffer overflows cannot corrupt any safe stack data.

Since the compiler must emit code to manipulate two stacks when enforcing SafeStack, this introduces modifications to both the prologue and epilogue of affected functions.  Thus, we include SafeStack as a distinct attack technique.  An example of SafeStack prologue and epilogue modifications is shown in Listing~\ref{lst:safestack}~(\S\ref{sec:inadvertent_evasion_effects}).

\mypara{Function Alignment}\label{par:alignment}%
Compilers provide a number of options to control the alignment of functions in memory.  Aligning functions to particular address boundaries can be advantageous from a performance perspective for architectural reasons, and optimal alignment varies depending on the target architecture.  On the other hand, compilers can also be instructed to eschew optional alignment constraints in favor of optimizing for size.  In this case, functions will be tightly packed and not conform to an alignment scheme.

The reason we include function alignment as an attack technique is two-fold.  First, tightly packing functions will remove any interstitial padding between adjacent functions, effectively creating a large change in instruction bytes preceding function prologues and succeeding function epilogues.  Second, varying the requested alignment will cause compilers to emit different sequences of padding instructions.  This leads to a similar, albeit weaker, change in prologue and epilogue-adjacent
instructions.  An example of this phenomenon is shown in Listing~\ref{lst:alignment}~(\S\ref{sec:inadvertent_evasion_effects}).

\subsection{Adversarial Attacks}%
\label{sub:adversarial_attacks}

\begin{listing}[t]
    \rule{\linewidth}{0.5pt}
\begin{minted}{nasm}
f_original:
    push rbp                ; save caller fp
    mov rbp, rsp            ; set fp
    ; function body...
    pop rbp                 ; restore caller fp
    ret                     ; return to caller
\end{minted}
    \rule{\linewidth}{0.5pt}
\begin{minted}{nasm}
f_injected:
    jmp .entry              ; jump over attack sequence
    mov eax, 0x89485590     ; attack sequence as immediate
.entry:
    nop                     ; prologue nop sequence tail
    push rbp                ; save caller fp
    mov rbp, rsp            ; set fp
    ; function body...
    pop rbp                 ; restore caller fp
    ret                     ; return to caller
    add word [rcx], al      ; adversarial insertion
    leave                   ; adversarial insertion
    ret                     ; adversarial insertion
    nop                     ; epilogue nop sequence tail
\end{minted}
    \rule{\linewidth}{0.5pt}
    \caption{Adversarial attack sequence injection example using compiler-emitted NOP sequences (additions in green).
    In the prologue, a relative jump is injected to bypass an instruction containing an attack sequence encoded as an immediate value.  In the epilogue, an attack sequence is directly injected verbatim; it will not be executed due to the unconditional return at line~10.
    }
    \label{lst:adversarial_sequences}
\end{listing}

In addition to the inadvertent attack techniques we just described, we also separately consider adversarial attack techniques.  Consistent with our two-tier threat model introduced in~\S\ref{sub:threat_model}, adversarial attacks go beyond the inadvertent evasive or false positive-inducing inputs that can be emitted by common compiler toolchains and configurations.  Instead, under this stronger threat model an adversary can use arbitrary techniques to craft a binary that will induce misclassifications by a function boundary detection NBA.

The possession of the power to arbitrarily modify binaries does not itself imply the ability to easily discover input byte and instruction sequences that produce misclassifications.  However, we find that an unguided search over bounded byte sequences is wholly sufficient to quickly find adversarial inputs that produce significant numbers of false positive or false negative misclassifications in the state of the art.

In particular, we explore the simple technique of injecting arbitrary byte sequences into function epilogues for this purpose.  In principle, one could use a binary rewriting framework~\cite{williams-king_2020_egalitolayoutagnosticbinary} to perform the injections on arbitrary in a functionality-preserving manner -- e.g., that makes the necessary modifications to account for the increased size of the code sections of the mutated binary.  However, we take the comparatively simpler approach of recompiling the binary corpus with a compiler configuration that causes NOP sequences of a desired length to be emitted in all epilogues.  This renders it straightforward to inject the necessary code to perform attack validation in a length-preserving manner via purely local modifications.

We employ and evaluate two forms of adversarial injection in terms of content:
\begin{enumerate*}[(i)]
    \item injecting a relative jump over a \texttt{mov} instruction that loads a register with the attack sequence as an immediate value, and
    \item injecting the attack sequence as-is into a function epilogue after a return instruction.
\end{enumerate*}
Due to the unconditional jump or return that prefaces each form of the injected attack sequence, there is no realistic possibility that the attack sequence will be executed in either form.  Listing~\ref{lst:adversarial_sequences} presents an example of this technique in action.  We note that while the injected code sequences could perhaps be identified as dead code and removed, the ability to do this reliably degrades quickly as more complicated instruction sequences are injected (up to the level of an opaque predicate).  We revisit this point in \S\ref{sec:discussion}; however, we do not believe it to be straightforward to identify and remove attack sequences injected by a determined adversary.

\section{Implementation}%
\label{sec:implementation}

The inadvertent attack search is implemented using an augmented version of BinKit~\cite{kim_2021_revisitingbinarycode}.  This framework provides scripts to reproducibly build a number of independent compiler toolchains (i.e., several versions of gcc and clang) as well as to download and compile numerous open source software packages using a variety of configurations.  We modified BinKit to support more compiler versions and configurations, and discuss the resulting experimental setup and data in \S\ref{sec:evaluation}.

Our adversarial attacks are implemented via a binary rewriting framework~\cite{agadakos_2019_nibblerdebloatingbinary} that in turn is based upon open source code drawn from pyelftools~\cite{bendersky_2012_pyelftools} and Capstone~\cite{capstonedevelopers_2014_capstonedisassembler}.  The framework operates on all ISAs present in BinKit.

We consider the binary rewriting procedure safe since it simply overwrites a number of NOP instructions placed in function epilogues by the compiler using the \texttt{-fpatch\-able-function-entry} option.  This preserves the existing binary layout in terms of addressing, and thus all jump and call targets remain valid.  Additionally, all injected code is protected by jumps or pre-existing return instructions that guard their execution.  Nevertheless, we manually spot-checked the rewriting procedure as well as ran existing test suites on modified binaries when available.

\section{Evaluation}%
\label{sec:evaluation}

In this section, we present the results of our evaluation of two representative state-of-the-art neural binary analyses for function boundary detection.  Our aim in conducting this evaluation was to answer the following research questions.

\newrq{inadvertent}
\newrq{adversarial}
\newrq{attacks}
\newrq{training}
\newrq{adtraining}

\begin{itemize}[noitemsep]
    \item[(\refrq{inadvertent})] Are NBAs susceptible to inadvertent attacks?
    \item[(\refrq{adversarial})] Are NBAs susceptible to adversarial attacks?
    \item[(\refrq{attacks})] Can an adversary systematically leverage inadvertent and adversarial attacks?
    \item[(\refrq{training})] Can inadvertent attacks be mitigated with larger, more representative training sets?
    \item[(\refrq{adtraining})] Can adversarial attacks be mitigated by including adversarial examples during training?
\end{itemize}

\subsection{Experimental Setup}%
\label{sub:experimental_setup}

\begin{table}
    \footnotesize
    \centering
    \caption{BinKit corpus.}%
    \label{table:dataset_overview}
    \setlength\tabcolsep{3.5pt}
    \begin{tabular}{lrrrrr}
        \toprule
        Dataset &   Binaries & Functions &  Packages &  Compilers &  Optimizations \\
        \midrule
        Normal       &     14,480 & 4,273,807 &        53 &         13 &              6 \\
        SizeOpt      &      2,115 &   575,143 &        51 &          9 &              1 \\
        NoInline     &      8,460 & 2,912,548 &        51 &          9 &              4 \\
        PIE          &      4,500 & 1,868,470 &        46 &          9 &              4 \\
        Obfuscate    &      4,700 & 1,351,779 &        51 &          4 &              5 \\
        CFI          &      3,800 & 1,133,310 &        52 &          3 &              5 \\
        ASE18        &     11,254 & 1,793,278 &         6 &         13 &              5 \\
        \bottomrule
    \end{tabular}
\end{table}

\begin{table}[h]
    \footnotesize
    \centering
    \caption{Results per dataset. For each dataset and metric (precision, recall, F1), the maximum value is highlighted green while the minimum value is highlighted red.  Large standard deviations (SD) are set in bold.}%
    \label{table:dataset_results}
    \begin{tabular}{llrrr}
        \toprule
                    &        & Precision      & Recall      & F1 \\
            Dataset & Tool   &      Mean (SD) &   Mean (SD) &  Mean (SD) \\
        \midrule
        Normal       & IDA     & \cellcolor{green!25} 1.000  (0.002) & \cellcolor{red!10}    0.844  \textbf{(0.144)} & \cellcolor{red!10}   0.908  (0.090) \\
                     & XDA     &                      0.989  (0.019) & \cellcolor{green!25}  0.965  (0.052) & \cellcolor{green!25} 0.976  (0.034) \\
                     & DeepDi  & \cellcolor{red!10}   0.976  (0.045) &                       0.932  (0.057) &                      0.952  (0.040) \\
        \midrule
        SizeOpt      & IDA     & \cellcolor{green!25} 1.000  (0.001) & \cellcolor{red!10}    0.754  \textbf{(0.155)} & \cellcolor{red!10}   0.851  \textbf{(0.103)} \\
                     & XDA     & \cellcolor{red!10}   0.977  (0.020) & \cellcolor{green!25}  0.900  (0.080) & \cellcolor{green!25} 0.935  (0.050) \\
                     & DeepDi  &                      0.987  (0.032) &                       0.870  (0.061) &                      0.923  (0.040) \\
        \midrule
        NoInline     & IDA     & \cellcolor{green!25} 1.000  (0.002) & \cellcolor{red!10}    0.848  \textbf{(0.156)} & \cellcolor{red!10}   0.909  (0.099) \\
                     & XDA     &                      0.991  (0.020) & \cellcolor{green!25}  0.954  (0.049) & \cellcolor{green!25} 0.971  (0.031) \\
                     & DeepDi  & \cellcolor{red!10}   0.980  (0.041) &                       0.945  (0.045) &                      0.961  (0.037) \\
        \midrule
        PIE          & IDA     & \cellcolor{green!25} 1.000  (0.003) & \cellcolor{red!10}    0.918  (0.097) &                      0.954  (0.059) \\
                     & XDA     &                      0.988  (0.022) & \cellcolor{green!25}  0.969  (0.050) & \cellcolor{green!25} 0.978  (0.034) \\
                     & DeepDi  & \cellcolor{red!10}   0.973  (0.058) &                       0.926  (0.064) & \cellcolor{red!10}   0.946  (0.053) \\
        \midrule
        Obfuscate    & IDA     & \cellcolor{green!25} 1.000  (0.001) & \cellcolor{red!10}    0.843  \textbf{(0.132)} & \cellcolor{red!10}   0.909  (0.082) \\
                     & XDA     & \cellcolor{red!10}   0.920  (0.085) & \cellcolor{green!25}  0.978  (0.034) &                      0.946  (0.050) \\
                     & DeepDi  &                      0.973  (0.040) &                       0.932  (0.055) & \cellcolor{green!25} 0.950  (0.034) \\
        \midrule
        CFI          & IDA     & \cellcolor{green!25} 1.000  (0.002) & \cellcolor{red!10}    0.806  \textbf{(0.183)} & \cellcolor{red!10}   0.880  \textbf{(0.122)} \\
                     & XDA     &                      0.975  (0.031) & \cellcolor{green!25}  0.883  \textbf{(0.111)} & \cellcolor{green!25} 0.923  (0.075) \\
                     & DeepDi  & \cellcolor{red!10}   0.968  (0.047) &                       0.880  (0.073) &                      0.919  (0.049) \\
        \midrule
        ASE18        & IDA     & \cellcolor{green!25} 1.000  (0.001) & \cellcolor{red!10}    0.832  \textbf{(0.117)} & \cellcolor{red!10}   0.904  (0.069) \\
                     & XDA     & \multicolumn{3}{c}{scores omitted: \textit{ASE18 was the training dataset}}                                      \\
                     & DeepDi  & \cellcolor{red!10}   0.977  (0.025) &                       0.957  (0.021) &                      0.966  (0.018) \\
        \midrule
        Totals       & IDA    & \cellcolor{green!25}  1.000 (0.002) & \cellcolor{red!10}     0.843 \textbf{(0.145)} & \cellcolor{red!10}    0.907  (0.092) \\
                     & XDA    &                       0.982 (0.040) & \cellcolor{green!25}   0.959 (0.062) & \cellcolor{green!25}  0.969  (0.044) \\
                     & DeepDi & \cellcolor{red!10}    0.976 (0.042) &                        0.931 (0.058) &                       0.951  (0.041) \\
        \bottomrule
    \end{tabular}
\end{table}

\mypara{Models Under Test}%
We selected the commercial standard IDA Pro~v7.7 as a baseline deterministic disassembler.  As exemplars of state-of-the-art neural binary analyses for function boundary detection, we selected XDA~\cite{pei_2021_xdaaccuraterobust} and DeepDi~\cite{yu_2022_deepdilearningrelational}, both of which we previously introduced in~\S\ref{sec:motivation}.  XDA was selected for evaluation because its design is heavily inspired by Transformer~\cite{vaswani_2017_attentionallyou}, and its implementation on top of Fairseq's implementation of BERT~\cite{devlin_2019_bertpretrainingdeep} reflects this.  As such, it is a perfect example of an NLP-based approach to neural function boundary detection.  DeepDi, on the other hand, was selected as an example of a function boundary detection system that incorporates some semantic information in the form of a graph model of instruction dependencies.  Finally, both of these systems publish a public artifact for evaluation: source code in the case of XDA~\cite{kexinpei_2021_xda}, and a binary distribution in the case of DeepDi~\cite{deepbitsdevelopers_2022_deepdi}.  We thank the authors of these systems for their commitment to open science.

\mypara{Datasets}%
To support reproducibility, we built our evaluation upon datasets generated for previous binary analysis evaluations.  In particular, we started with the BinKit corpus~\cite{kim_2021_revisitingbinarycode} which is based on all available GNU software packages.  BinKit includes 53 software packages compiled by five versions of GCC (v4.9.4, v5.5.0, v6.4.0, v7.3.0, v8.2.0) and four versions of clang (v4.0, v5.0, v6.0, v7.0).  The original corpus is composed of several distinct datasets that exercise
specific compiler options: \texttt{-fno-inline}, \texttt{-fPIE}, \texttt{-Os}, and \texttt{-flto}.  We expanded the corpus to include more recent versions of GCC and Clang (GCC~v.9.4.0, GCC~v11.2.0, Clang~v9.0, Clang~v13.0) and a new dataset (CFI).  The CFI dataset exercises modern control-flow integrity compiler options discussed in \S\ref{par:cfi} via the GCC compiler flag \texttt{-fcf-protection=full} and the Clang compiler flag \texttt{-fsani\-tize=cfi}.  An overview of the corpus and individual datasets is presented in Table~\ref{table:dataset_overview}.

Although the BinKit corpus includes a substantial combination of compiler versions, optimization levels, and specific flags, one cannot assume that compiler options are completely isolated for any particular binary or dataset.  For example, one might assume the NoInline dataset would not include code that had been compiled with the flag \texttt{-fgnu89-inline}, which causes inlining, or that a binary compiled at the \texttt{O0} optimization level would not include code compiled at a different optimization level.  Unfortunately, this is not the case due to existence of compiler-generated code and code that is statically linked in from compiler support libraries.  We found that binaries compiled with \texttt{-fstack-protector-strong} included code compiled with \texttt{-fno-stack-protector}, although the presence of the latter was dominated by the former.  In some cases, such as the xorriso binary with \num{3000} functions, compiler support code is dominated by the software library code, and thus the presence of code compiled with different flags would have a minor impact on training and evaluation.  On the other hand, a software library like coreutils is composed of many small utilities where the ratio of compiler support code to library code is much less.  We do not believe that this phenomenon has a significant impact on our results, but we do note that it is non-trivial to ensure uniform compiler configurations on absolutely all code in each dataset and that we did not attempt to achieve this.

\mypara{Metrics}%
We report precision, recall, and the balanced F-score ($F_1$ score) with the standard definitions.  In Table~\ref{table:dataset_results}, we report the mean and standard deviation of the precision, recall, and $F_1$ score as a statistical summary calculated per binary in each dataset.  We choose to report mean and standard deviation because performance within a particular dataset can exhibit high variance, as we discuss later.

\mypara{Computational Resources}%
All experiments were performed on a dedicated server with a 64 core AMD Ryzen 3995WX CPU @ 4.3GHz, three RTX A6000 GPUs, 1TB memory, and a 4TB SSD.

\subsection{\refrq{inadvertent}: Inadvertent Attacks}%
\label{sub:rq_inadvertent}

\begin{figure*}[t]
  \includegraphics[width=1.0\linewidth]{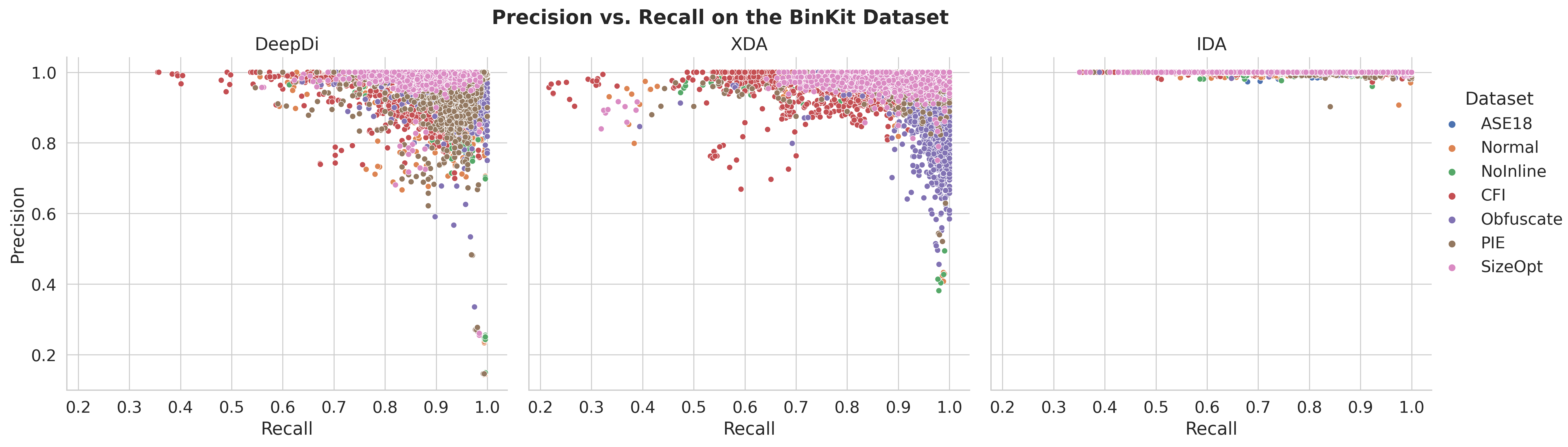}
  \caption{Overview of precision versus recall per binary from the BinKit corpus. IDA Pro consistently performs best with respect to precision, with little variance. XDA, however, dominates with respect to recall. It also wins out on F1 score in all but one case, Obfuscate, where DeepDi is best. Variance in these metrics is somewhat apparent, but better observed in Figure~\ref{fig:adversarial_attack_hist}.}%
  \label{fig:binkit-precision-vs-recall}
\end{figure*}

In our first experiment, we subjected XDA and DeepDi to our augmented version of BinKit to evaluate their resilience to the full set of inadvertent evasions described in~\S\ref{sub:inadvertent_attacks}.  We additionally include IDA Pro in this experiment as a baseline representing the state of the art in deterministic function boundary detection.  Table~\ref{table:dataset_results} presents summary statistics in terms of precision, recall, and F1 score, along with standard deviation for each metric, broken out by the individual datasets comprising BinKit.

From the data, IDA Pro consistently performs best with respect to precision, with little variance.
XDA, however, dominates with respect to recall and F1 score.
DeepDi produces F1 scores that are very close to the performance of XDA and takes the top spot for exactly one dataset, Obfuscate.
There is also clearly some variance across all metrics.
However, in this respect the summary statistics do not tell the full story.

Figure~\ref{fig:binkit-precision-vs-recall} presents a series of precision-recall plots for each system.  Each point represents one binary, colored according to membership in each of BinKit's constituent datasets.  In each plot, the optimal point is the upper-right, indicating perfect precision (all detections were true positives) and recall (all functions were detected).  Points towards the x-axis indicate lower precision and thus a higher proportion of false positives.  Points towards the y-axis (left) indicate lower recall and thus a higher proportion of false negatives.

One can immediately observe a marked difference between the operating characteristics of the deterministic baseline represented by IDA Pro and the NBA systems.  IDA Pro consistently achieves near perfect precision -- i.e., when it detects a function, it is highly likely to be a true positive.  However, it is prone in some cases to unreported functions.  In the worst case, IDA Pro dips below \num{0.4} recall.

Both XDA and DeepDi, however, exhibit much stronger variance in both precision and recall.
XDA in particular presents seeming clusters, i.e., precision-recall that correlates with individual datasets.
XDA performs particularly poorly on the CFI dataset, colored in red.
Other datasets are biased towards either precision or recall failures.
For instance, Obfuscate, colored in purple, tends towards lower recall and false negatives.
SizeOpt failures, in contrast, are biased towards lower precision and false positives.

In comparative terms, XDA performs slightly better across the board than XDA and both exhibit better recall than IDA Pro on this data.  However, the scatterplot makes it clear that there is a sizable number of outliers in both precision and recall.  Thus, we investigated a sample of these outliers.

\begin{listing}[h]
\begin{minted}{nasm}
sub    rsp, 0x8         ; align stack pointer
push   r13              ; push arg7
mov    r9, r12          ; set arg6
mov    r8, rbp          ; set arg5
mov    ecx, 0x4972ec    ; set arg4
mov    rdx, rbx         ; set arg3
mov    rsi, rax         ; set arg2
mov    rdi, stderr      ; set arg1
mov    eax, 0x0         ; zero rax
call   fprintf          ; invoke fprintf
\end{minted}
    \caption{One example of a single instruction that causes DeepDi to issue >\num{3000} false positives for the gcal-4.1 benchmark.}
    \label{lst:gcal}
\end{listing}

One such outlier point is shown in Listing~\ref{lst:gcal}.  On the gcal-4.1 benchmark, DeepDi issued >\num{3000} distinct false positives from multiple occurrences of a single instruction.  The instruction, highlighted in red, subtracts \num{8} bytes from the stack pointer.  This is an operation that is often performed in a function prologue to allocate space for local variables on the stack.  However, this particular example occurs when marshalling arguments to a call to \texttt{fprintf} in gcal's \texttt{main} function.  The reason this occurs is because this particular call to \texttt{fprintf}, a variadic function, has more than six arguments.  The SysV ABI dictates that the first six arguments are passed in registers, while any further arguments are passed on the stack.  Stack arguments, however, must be aligned to a 16-byte boundary.  This causes the compiler, which was configured to operate at \texttt{O0} in this case, to directly adjust the stack pointer prior to pushing the seventh argument to \texttt{fprintf}.  It appears that the DeepDi model we evaluated never observed this particular pattern in its training set.


One can argue that if the failures we observe are restricted to accidental outliers, then their overall impact should be low.  Unfortunately, as we demonstrate next, these inadvertent misclassifications can be systematically exploited by an adversary to build effective adversarial attacks.

\subsection{\refrq{adversarial}: Adversarial Attacks}
\label{sub:rq_adversarial}

\begin{figure}
    \centering
    \includegraphics[width=1.0\columnwidth]{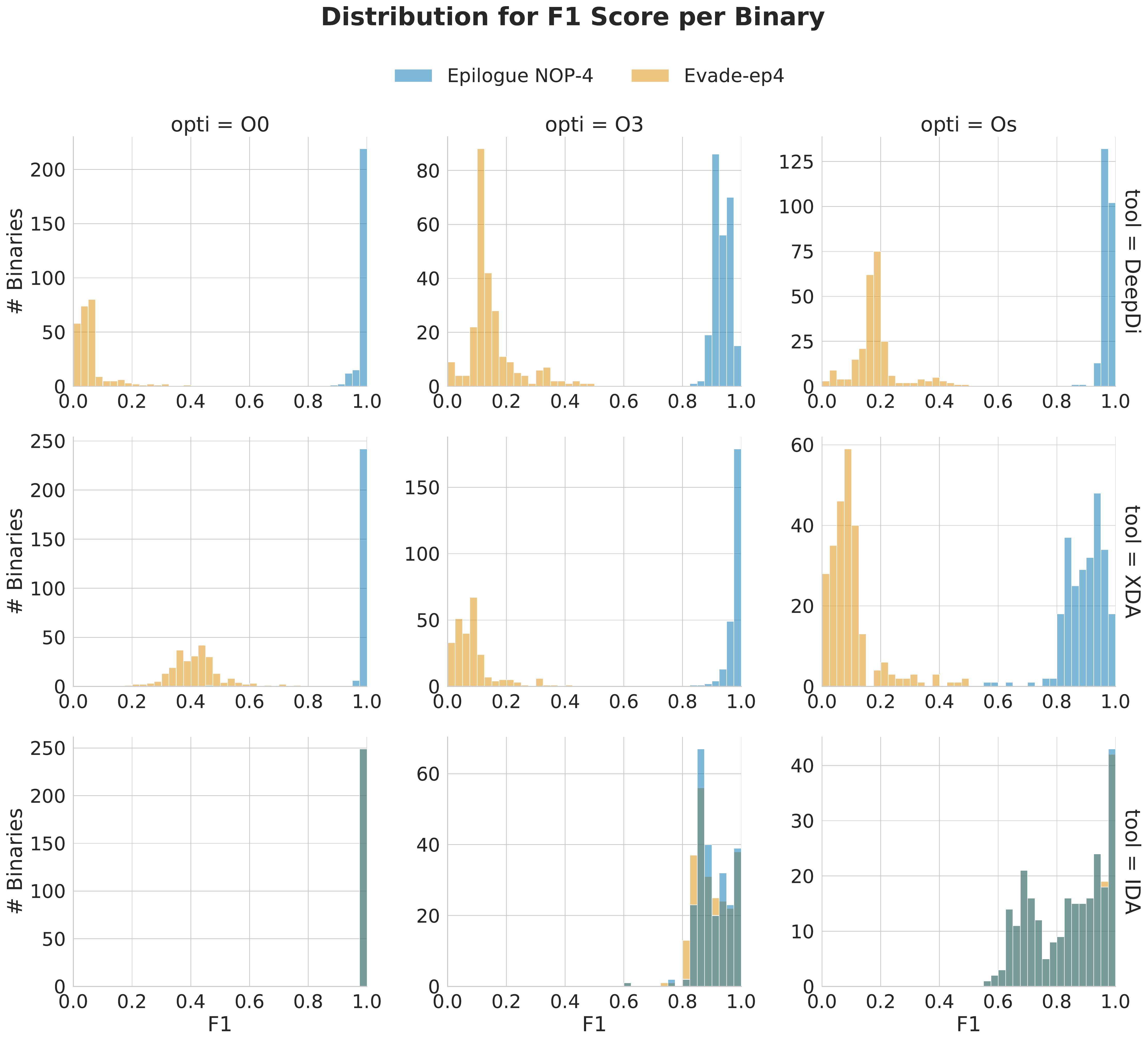}
    \caption{Effects of adversarial attacks on F1 score.
    We recompile the Normal dataset with the compiler option \texttt{-fpatchable-function-entry=4,4}, inserting 4 NOP instructions after each function under different optimization levels.
    Both XDA and DeepDi are resilient to the simple addition of NOP sequences as shown in the baseline experiments.  However, F1 scores exhibit significant degradation when injecting adversarial patterns into the NOP regions. IDA Pro is unaffected by epilogue mutation evasions.}%
    \label{fig:adversarial_attack_hist}
\end{figure}

To evaluate adversarial attack efficacy, we recompiled the Normal dataset with different optimization options (\texttt{O0}, \texttt{O3}, \texttt{Os}) and the \texttt{-fpatch\-able-function-entry=4,4} flag which inserts \num{4} NOP instructions after the original function epilogue.  The effects of this on F1 score are presented in Figure~\ref{fig:adversarial_attack_hist}.

Both XDA and DeepDi successfully handle the addition of simple ``NOP sled'' insertion, preserving high F1 scores. Unfortunately, when adversarial mutations are introduced following the methodology described in \S\ref{sub:adversarial_attacks}, both systems diverge significantly from their published accuracy.  Interestingly, we observe that XDA is more resilient to epilogue mutation under the \texttt{O0} optimization level versus \texttt{O3} and \texttt{Os}.  DeepDi's performance is degraded for all optimization levels, with median F1 scores well below 0.25 at all optimization levels.  IDA Pro, however, is largely unaffected by epilogue mutations as evidenced by the near-identical F1 distributions across both datasets.

\subsection{\refrq{attacks}: Systematic Attacks}%
\label{sub:rq_attacks}

Our results to this point highlight that both XDA and DeepDi are vulnerable to seemingly simple adversarial byte sequence injection, causing them to misclassify significant portions of the functions present using the same sequence across all binaries with no attempt to adapt them to a given program.  Unfortunately, during our evaluation we unearthed several cases where the patterns used were particularly effective, leading to almost complete evasion.  Specifically, XDA only managed to recover
\num{36} out of the \num{145} ground truth functions contained in \texttt{gnudos-1.11.4\_gcc-9.4.0\_\-x86\_64\_\-Os\_prime.elf}.  As another example, DeepDi reported only four (4) out of a total of \textbf{6,679} ground truth functions present in \texttt{gsl-2.5\_\-gcc-9.4.0\_\-x86\_64\_O0\_libgsl.so.23.1.0.elf}.  In the other direction, DeepDi reported reported \textbf{\num{3193}} out of the \num{1104} ground truth functions present in \texttt{gcal-4.1\_gcc-8.2.0\_\-x86\_64\-\_O1\_gcal.elf}.

We believe that these cases are due to these particular adversarial patterns being especially effective on the characteristic layout of those programs.  Furthermore, our adversarial attacks could potentially be improved by targeting them for particular binaries, paving the way for a novel, insidious way to attack NBAs relying only on static information.  As is evident in the DeepDi case, such adversarially mutated binaries would be virtually invisible to detectors that relied on a vulnerable NBA for as part of its analysis pipeline.  While the targeted attacks we speculate about here are beyond the scope of this work, we believe that it is a promising line of inquiry and plan to explore it as future work.

\subsection{\refrq{training}: Expanded Training Sets}%
\label{sub:rq_training}

To investigate whether inadvertent attacks can be mitigated with additional training, we next conducted a step-wise experiment with XDA.\footnote{We are restricted to XDA for this and the following experiment since DeepDi is distributed as a binary object.  Thus, we do not have the ability to train a new DeepDi model.}
For the inadvertent samples, we chose to evaluate XDA with the CFI dataset as it was the most difficult dataset to classify for all three systems under evaluation.
Starting with a very limited subset of the ASE18 dataset, we trained XDA with increasingly more diversity in the number of compilers, compiler versions, and compiler options.
The results are shown in Table~\ref{table:training_xda}.
With only one version of GCC and four optimization levels in the training data, XDA achieved a reasonable F1 score of 0.855 on the CFI dataset.
By adding four versions of Clang and GCC, a modest improvement in F1 score obtained (0.865).
Notably, adding Clang-compiled binaries to the training set reduced XDA's performance.

\begin{table}
    \footnotesize
    \centering
    \caption{Improving resilience through training.}%
    \label{table:training_xda}
    \setlength\tabcolsep{3.5pt}
    \begin{tabular}{rlrrrrclr}
        \toprule
           &               &          & Data       &       & \multicolumn{2}{c}{\# GCC,}  & Eval    & \\
        ID & Base (+added) & \# Files & Size       & Time  & \multicolumn{2}{c}{\# Clang} & Dataset & F1 \\
        \midrule
         0 & GCC-6.4.0    & 772      & 118M       & 0.4h  & 1  &           & CFI      & 0.855  \\
         1 & Clang        & 3,088    & 461M       & 1.7h  &    &  4        & CFI      & 0.575  \\
         2 & GCC          & 3,860    & 586M       & 2.2h  & 5  &           & CFI      & 0.857  \\
         3 & original     & 6,948    & 1,046M     & 3.9h  & 5  &  4        & CFI      & 0.865  \\
         4 & 3 +Clang-new & 6,988    & 1,153M     & 4.3h  & 5  &  6        & CFI      & 0.852  \\
         5 & 3 +GCC-new   & 6,988    & 1,148M     & 4.3h  & 7  &  4        & CFI      & 0.913  \\
         6 & 3 +both-new  & 7,028    & 1,256M     & 4.7h  & 7  &  6        & CFI      & 0.923  \\
         7 & 6 +Os        & 7,058    & 1,312M     & 4.9h  & 7  &  6        & CFI      & 0.924  \\
        \midrule
         8 & 7 +nop4      & 7,133    & 1,499M     & 5.6h  & 7  &  6        & CFI      & 0.935  \\
         9 & 8 +evade4    & 7,208    & 1,685M     & 6.4h  & 7  &  6        & CFI      & 0.810  \\
        \midrule
         7 & 6 +Os        &          &            &       &    &           & Evade-ep4 & 0.198 \\
         8 & 7 +nop4      &          &            &       &    &           & Evade-ep4 & 0.338 \\
         9 & 8 +evade4    &          &            &       &    &           & Evade-ep4 & 0.938 \\
        \bottomrule
    \end{tabular}
\end{table}

We then expanded the original ASE18 dataset by including newer compilers, namely two versions of Clang and GCC, which increased the F1 score to 0.923.
Finally, by adding the \texttt{Os} optimization level, XDA achieved a score of 0.924, which is better than both DeepDi and IDA Pro.
This demonstrates that XDA's performance can in fact be improved by expanding the training dataset -- which is expected -- but also that XDA is also quite sensitive to compiler versions and options present in the training data.

\subsection{\refrq{adtraining}: Adversarial Training}%
\label{sub:rq_adtraining}

In the final experiment, we evaluate whether MUTs can be made resilient to the adversarial attacks we describe by adopting adversarial training.  In order to train XDA on these crafted attacks, we created a new dataset based on the NOP dataset described in \S\ref{sub:rq_adversarial}.
In this dataset, we replaced each 4-byte NOP epilogue with a randomly chosen evasion pattern that is also a valid 4-byte x86 instruction sequence.
We then fine-tuned XDA on this expanded dataset and evaluated on both the CFI and Evade-ep4 datasets.
With the new model, XDA's performance on the Evade-ep4 dataset improved from 0.198 to 0.938, a significant improvement.
Unfortunately, XDA's performance on the CFI dataset was also degraded from 0.924 to 0.810.  This suggests that while adversarial training can partially mitigate evasion, it also comes at a significant cost in accuracy for benign samples.

In addition, it is also unclear whether training on adversarial examples represents a trustworthy mitigation.  To illustrate, we performed an additional round of adversarial attack search to demonstrate the inherent limitation of training against adversarial techniques.
Repeating our 4-byte evasion search, we were able to reduce XDA's performance to 0.488 (STD 0.317) when trained on the Evade-ep4 dataset.
Additionally, we studied two alternative attacks using a 3-byte and 8-byte NOP dataset, producing F1 scores of 0.427 and 0.430 respectively.
Thus, while one would hope that training on adversarial examples would produce a model that is robust against many different evasion patterns, our experiments show that this is unlikely to be the case as we were able to degrade XDA's performance again without significant effort.

\section{Discussion}%
\label{sec:discussion}

\textbf{Black-box attacks are powerful enough.}  As is hopefully clear from our evaluation, black-box attacks are sufficiently powerful to discover numerous false positive and false negative-inducing inputs to current generation function boundary detection NBAs.  Sophisticated white-box searches for adversarial examples that rely on gradient descent might well find more attacks.  However, it is unclear how one might adapt existing searches while preserving the functionality of the mutated binary due to the discrete problem space.  Nevertheless, this is an interesting direction to explore.

\textbf{Inadvertent attacks break pure NLP-based systems.}  As should also be clear from the evaluation, inadvertent attacks significantly degrade function boundary detection approaches that directly reuse NLP embeddings and models as XDA does.  Another way to view this finding is that such approaches do not generalize well to examples that are not observed during training.  In retrospect, this naturally follows from our conjecture that syntactic representations are not a sound basis for binary analysis where semantics is virtually always what actually matters.  One could argue that simply including misclassified examples in the training set is sufficient mitigation, and there is likely some truth to that.  However, in our opinion a realistic counterargument is that anticipating and training on a sufficiently large permutation of compilers, compiler versions, and compiler configurations is combinatorially difficult.  To make matters worse, that mitigation does not take adversarial attacks into account.

\textbf{Domain-specific embeddings and graph models are a marginal improvement.}  The evaluation shows that DeepDi's domain-specific embedding and use of R-GCN to model instruction dependencies improves its resilience to inadvertent attacks.  This is clear evidence that incorporating even a small bit of the latent semantic information present in an instruction stream has utility.  However, this improvement is tempered by DeepDi's performance against adversarial examples, motivating our next observation.

\textbf{Focus on semantics instead of syntax.}  The overarching conclusion we draw from the evaluation is that syntactic representations are unlikely to be a reliable basis for binary analyses.  In a way, this is unsurprising, since syntactic approaches for attack detection such as signature-based IDS and first-generation anti-virus based on pattern matching against byte sequences were criticized for similar deficiencies long ago.  While these techniques can of course be useful, they cannot be relied upon in isolation.  Instead, mirroring attack detection's move from static pattern matching to dynamic behavioral analysis more than a decade ago, we argue that future work in this space should emphasize semantics over syntax to avoid similar pitfalls.

\textbf{Evaluation quality is important.}  In tandem with the semantics question, we believe it is crucial that the research community hews to a standard of evaluations on large, representative, public datasets. This data should include a range of programs with varying functionality, as well as different compilers, compiler versions, and compiler configurations. As shown in our experiments, testing on a more comprehensive dataset such as BinKit~\cite{kim_2021_revisitingbinarycode} over smaller, less representative datasets in the original papers can help identify areas of improvement for underlying models, such as lacking understanding of semantic isomorphisms. Finally, we believe that these benchmark corpora should include adversarial examples generated using techniques such as those described herein to directly test whether future work is susceptible to similar attacks. This inclusion should both to increase robustness in possible security related use cases and to help the model learn patterns of adversarial perturbation that exploit syntactic versus semantic model understanding. A substantial bonus in following such a standard would be to ease reproducibility and comparative evaluation.

\textbf{Detecting adversarial code is not easy.}  Finally, we readily acknowledge that the adversarial code we inject as part of our methodology and evaluation is likely to be easy to detect and strip before performing classification.  However, we believe that focusing on this is misguided.  Code obfuscation is well within the threat model of many contexts in which binary analyses such as function boundary detection operate under.  In that light, it is reasonable to suspect that if an adversary wished to do so, they could easily obfuscate the fact that the injected code will never be executed by relying on computed control transfers and opaque predicates.  Indeed, if a defender was able to perfectly identify dead code, then a large part of the debloating problem would be perfectly solved which -- to our knowledge -- is not the case.  Instead, as with so many other problems in this space, detecting and removing adversarial code reduces to Rice's Theorem~\cite{rice_1953_classes}.  Thus, we believe it is safe to conclude that this is not likely to be a fruitful research direction.

\section{Related Work}%
\label{sec:related_work}

\mypara{Neural binary analysis}%
Binary analysis is a long-studied and expansive research area.
Disassembly is a fundamental task that traditionally has been solved using deterministic algorithms that can be broadly classified as either linear disassembly (provided by tools like objdump from GNU binutils) or recursive descent disassembly (provided by tools such as IDA Pro~\cite{guilfanov_2022_idapro}, Ghidra~\cite{nsa_2019_ghidra}, or Binary Ninja~\cite{vector35_2016_binaryninja}).  These tools typically also incorporate algorithms for function boundary detection using some combination of symbol table information, debug information, and pattern-based heuristics.  Work such as ByteWeight~\cite{bao_2014_byteweightlearningrecognize} specifically investigated learning-based approaches for performing function boundary detection.  Other common binary analysis tasks include measuring similarity between snippets of binary code~\cite{haq_2019_surveybinarycode}, recovering source code types~\cite{lee_2011_tieprincipledreverse}, and decompilation~\cite{schwartz_2013_nativex86decompilation,yakdan_2015_nomoregotos}.
In recent years, applying deep learning techniques to binary analysis problems has become a popular topic of study due to the success of DNNs in solving image and text processing tasks, among others.  Shin et al.~\cite{shin_2015_recognizingfunctionsbinaries} were the first to apply a DNN to a binary analysis problem; in this case, detecting function boundaries using a bi-directional recurrent neural network (BiRNN).  The strategy of repurposing embeddings and model architectures originally developed to solve NLP or image processing problems became \emph{de rigeur} in a way.  Numerous NBAs for disassembly~\cite{pei_2021_xdaaccuraterobust,yu_2022_deepdilearningrelational}, function boundary detection~\cite{chua_2017_neuralnetscan,pei_2021_xdaaccuraterobust,yu_2022_deepdilearningrelational}, value set analysis~\cite{guo_2019_deepvsafacilitatingvalueset,li_2021_palmtreelearningassembly}, static code similarity~\cite{xu_2017_neuralnetworkbasedgraph,ding_2019_asm2vecboostingstatic,zuo_2019_neuralmachinetranslation,lee_2019_instruction2vecefficientpreprocessor,massarelli_2018_safeselfattentivefunction,yu_2020_ordermatterssemanticaware,li_2021_palmtreelearningassembly,yang_2021_codeetensorembedding}, decompilation~\cite{fu_2019_codaendtoendneural}, and malware analysis~\cite{huang_2016_mtnetmultitaskneural,al-dujaili_2018_adversarialdeeplearning,vinayakumar_2019_robustintelligentmalware} directly use embeddings~(e.g., word2vec~\cite{mikolov_2013_efficientestimationword}, PV-DM~\cite{le_2014_distributedrepresentationssentences}) or models~(e.g, RNN, CNN, Transformer~\cite{vaswani_2017_attentionallyou}, BERT~\cite{devlin_2019_bertpretrainingdeep}) developed for the NLP or image problem domains.  One of the conclusions we draw in this paper is that while it is tempting to build on techniques that have been successful in other areas, binary analysis is a strikingly different research area with a different threat model and much stronger accuracy requirements for downstream tasks~(see the discussion in~\S\ref{sec:discussion}).  For NBAs to be resilient against adversaries that seek to evade or confuse binary analyses, choices of embeddings and model architectures should reflect these requirements.

We are not the first to independently evaluate NBA systems for other tasks.  Kim et al.~\cite{kim_2021_revisitingbinarycode} studied NBAs that perform static similarity detection using a large dataset of programs compiled with a variety of toolchains and compiler options called BinKit; we build on BinKit to carry out our own evaluation.  Using this dataset and a simple baseline similarity detector called TikNib, they show that NBAs do not necessarily outperform simpler, explainable methods such as the one implemented by TikNib.  Marcelli et al.~\cite{marcelli_2022_howmachinelearning} performed a similar study also focused on static similarity detection NBAs, and show that published results do not necessarily hold when the systems-under-test are trained and evaluated on larger, more representative datasets.  Finally, Lucas et al.~showed that DNNs used for static malware detection on binary programs are prone to adversarial attacks~\cite{lucas_2021_malwaremakeoverbreaking}.  This work lies in contrast to our own not only in the specific problem domain but also in its use of traditional adversarial ML techniques -- i.e., white-box gradient descent or black-box hill climbing -- to find evading transformations.

\mypara{Adversarial machine learning}%
Substantial research has studied the problem of crafting adversarial examples~\cite{biggio_2018_wild, papernot_2018_aml_sok}. Traditionally, this research has been conducted on semi-continuous spaces, here defined as when adjacent values carry semantic information, e.g., pixel values for image classification. In these approaches, attacks use a variety of derivative-based approaches to optimize loss over some non-convex objective function~\cite{guo_2019_simple, moosavi_2016_deepfool, goodfellow_2014_explaining, athalye_2018_obfuscated, carlini_2017_towards, biggio_2013_evasion, szegedy_2013_intriguing}. In our case, we examine executable binaries, where we must work under more difficult constraints. First, adjacent values for binary code do not carry semantic meaning. For instance, 0x8F is the binary encoding of the x86 \texttt{pop} instruction, whereas 0x90 is the semantically unrelated \texttt{nop} instruction. This difference is non-trivial as it presents a much harder problem than that of optimization over semi-continuous spaces; in fact, it reduces to integer factorization, an NP-complete problem~\cite{karp_1972}. Pierazzi et al.~\cite{pierazzi_2020_intriguingpropertiesadversarial} provide detailed insight into how different problem spaces under which adversarial machine learning is conducted, such as using binary code as the input to a DNN, require specific black-box attacks because traditional gradient-based approaches fail. Another constraint we must satisfy is to produce valid executable binaries. These constraints are similar to those necessary in any attack that attempts to modify binary code~\cite{lucas_2021_malwaremakeoverbreaking, anderson_2018_learning_to_evade}.

As stated in the previous subsection, other work in using deep learning for malware analysis has looked the problem of mapping binaries to either malicious or benign software~\cite{raff_2017_malconv,raff_2018_malwaredetectioneating}. In turn, various work has aimed to attack this type of machine learning model and others like it~\cite{lucas_2021_malwaremakeoverbreaking, anderson_2018_learning_to_evade}. However, this paper presents the first exploration into evaluating the robustness of deep learning models against both inadvertent attacks and crafted adversarial examples.

\section{Conclusions and Future Work}%
\label{sec:conclusions}

We presented the first study of the resilience of neural function boundary detectors to inadvertent and adversarial attacks.  Our methodology demonstrates that straightforward black-box search using a large dataset and toolchain array is sufficient to identify numerous adversarial examples for two state-of-the-art systems: XDA~\cite{pei_2021_xdaaccuraterobust} and DeepDi~\cite{yu_2022_deepdilearningrelational}; sophisticated white-box search algorithms are unnecessary.  Our conjecture -- which we believe is validated by our evaluation -- is that these systems are susceptible to attack because they rely on embeddings and model architectures intended for syntactic inference, and do not sufficiently consider the semantics of the ISAs they operate on.

This is not to say that this research direction should be abandoned.  To the contrary, we believe there remains significant potential for applying deep learning to binary analysis problems.  However, future research might well benefit from focusing on instruction semantics rather than syntactic representations.  In addition, future work should ensure that evaluations are based on large, representative datasets that includes adversarial examples intended to exploit syntactic dependence.  An intriguing research question is whether effective embeddings and model architectures can be developed specifically for binary analysis tasks.  We plan to investigate this question in our future work, and hope others will as well.

\begin{acks}
\scriptsize{%
\noindent
The views expressed herein are those of the author and do not reflect the position of the United States Military Academy, the Department of the Army, or the Department of Defense.%

This material is based upon work supported by the National Science Foundation under Grant No.\ 2203175.%

}
\end{acks}

\bibliographystyle{ACM-Reference-Format}
\bibliography{bib/extracted}

\newpage
\appendix

\section{Inadvertent Evasion Effects}%
\label{sec:inadvertent_evasion_effects}

\begin{listing}[H]
    \rule{\linewidth}{0.5pt}
\begin{minted}{nasm}
f_no_stack_protector:
    push   rbp          ; save caller frame pointer
    mov    rbp, rsp     ; set frame pointer
    ; function body...
    pop    rbp          ; restore caller frame pointer
    ret                 ; return to caller
\end{minted}
    \rule{\linewidth}{0.5pt}
\begin{minted}{nasm}
f_stack_protector_strong:
    push   rbp                      ; save caller fp
    mov    rbp, rsp                 ; set fp
    sub    rsp, 0x10                ; alloc stack
    mov    rax, qword fs:0x28       ; get global guard
    mov    qword [rbp-0x8], rax     ; inject guard on stack
    ; function body...
    mov    rax, qword fs:0x28       ; get global guard
    mov    rcx, qword [rbp-0x8]     ; get stack guard
    cmp    rax, rcx                 ; compare guards
    jne    .fail                    ; jump if not equal
    add    rsp, 0x10                ; dealloc stack
    pop    rbp                      ; restore fp
    ret                             ; return to caller
.fail:
    call   __stack_chk_fail         ; terminate program
\end{minted}
    \rule{\linewidth}{0.5pt}
    \caption{Stack protector function prologue and epilogue modifications (additions in green).}
    \label{lst:stack_protector}
\end{listing}

\begin{listing}[H]
    \rule{\linewidth}{0.5pt}
\begin{minted}{nasm}
f_no_stack_clash_protection:
    push   rbp              ; save caller fp
    mov    rbp, rsp         ; set fp
    sub    rsp, 0x10020     ; allocate buffer
    ; function body...
\end{minted}
    \rule{\linewidth}{0.5pt}
\begin{minted}{nasm}
f_stack_clash_protection:
    push   rbp                  ; save caller fp
    mov    rbp, rsp             ; set fp
    mov    r11, rsp             ; get sp
    sub    r11, 0x10000         ; set stack alloc size
.next_page:
    sub    rsp, 0x1000          ; alloc next stack page
    mov    qword [rsp], 0x0     ; probe page with store
    cmp    rsp, r11             ; check if done allocing
    jne    .next_page           ; if not, loop
    sub    rsp, 0x20            ; alloc final 0x20
    ; function body...
\end{minted}
    \rule{\linewidth}{0.5pt}
    \caption{Stack clash protection function prologue modifications (deletions in red, additions in green).}
    \label{lst:stack_clash}
\end{listing}

\begin{listing}[H]
    \rule{\linewidth}{0.5pt}
\begin{minted}{nasm}
f_no_cet:
    push   rbp              ; save caller fp
    mov    rbp, rsp         ; set fp
    sub    rsp, 0x10        ; alloc stack
    ; function body...
\end{minted}
    \rule{\linewidth}{0.5pt}
\begin{minted}{nasm}
f_cet:
    endbr64                 ; label valid branch target
    push   rbp              ; save caller fp
    mov    rbp, rsp         ; set fp
    sub    rsp, 0x10        ; alloc stack
    ; function body...
\end{minted}
    \rule{\linewidth}{0.5pt}
    \caption{Intel CET function prologue modifications (additions in green).}
    \label{lst:cfi}
\end{listing}

\begin{listing}[H]
    \rule{\linewidth}{0.5pt}
\begin{minted}{nasm}
f_no_safe_stack:
    push   rbp              ; save caller fp
    mov    rbp, rsp         ; set fp
    sub    rsp, 0x1010      ; alloc stack vars
    ; function body...
    add    rsp, 0x1010      ; dealloc stack vars
    pop    rbp              ; restore caller fp
    ret                     ; return

\end{minted}
    \rule{\linewidth}{0.5pt}
\begin{minted}{nasm}
f_safe_stack:
    push   rbp              ; save caller fp
    mov    rbp, rsp         ; set fp
    sub    rsp, 0x20        ; alloc safe stack
                            ; get unsafe stack pointer
    mov    rcx, __safestack_unsafe_stack_ptr
    ; function body...
    add    rsp, 0x20        ; dealloc safe stack
    pop    rbp              ; restore caller fp
    ret                     ; return

\end{minted}
    \rule{\linewidth}{0.5pt}
    \caption{SafeStack function prologue and epilogue modifications (deletions in red, additions in green).}
    \label{lst:safestack}
\end{listing}

\begin{listing}[H]
    \rule{\linewidth}{0.5pt}
\begin{minted}{nasm}
    add    rsp, 0x8         ; prev fn epilogue
    pop    rbx
    pop    r14
    ret
f_unaligned:
    push   rbp              ; save caller registers
    push   r15
    push   r14
    push   r13
    push   r12
    push   rbx
    sub    rsp, 0xe8        ; alloc stack vars
    ; function body...
\end{minted}
    \rule{\linewidth}{0.5pt}
\begin{minted}{nasm}
    add    rsp, 0x8         ; prev fn epilogue
    pop    rbx
    pop    r14
    ret
    int3                    ; sw bkpt padding bytes
    int3
    int3
    ; many repetitions...
    int3
    int3
    int3
f_aligned:
    push   rbp              ; save caller registers
    push   r15
    push   r14
    push   r13
    push   r12
    push   rbx
    sub    rsp, 0xe8        ; alloc stack vars
    ; function body...
\end{minted}
    \rule{\linewidth}{0.5pt}
    \caption{Interstitial function padding modifications due to varying alignment constraints (additions in green).}
    \label{lst:alignment}
\end{listing}

\end{document}